\documentclass[reprint, prd]{revtex4-1}

\usepackage[utf8]{inputenc}
\usepackage{amsmath}
\usepackage{amssymb}
\usepackage{mathrsfs}
\usepackage{times}
\usepackage{hyperref}
\usepackage{natbib}
\usepackage{graphicx}
\usepackage{hyperref}
\usepackage{natbib}

\hypersetup{
  bookmarksnumbered = true,
  bookmarksopen=false,
  pdfborder=0 0 0,         % make all links invisible, so the pdf looks good when printed
  pdffitwindow=true,      % window fit to page when opened
  pdfnewwindow=true, % links in new window
  colorlinks=true,           % false: boxed links; true: colored links
  linkcolor=blue,            % color of internal links
  citecolor=magenta,    % color of links to bibliography
  filecolor=magenta,     % color of file links
  urlcolor=cyan              % color of external links
}

\newcommand{\msun}{\ensuremath{M_\odot\,}}
\newcommand{\chimera}{{\sc Chimera }}

\newcommand{\gcc}{\ensuremath{{\mbox{g~cm}}^{-3}}}

\newcommand{\apjl}{Astrophys. J. Lett.}
\newcommand{\mnras}{Mon. Not. R. Astron. Soc.}

\begin{document}

\title{The Gravitational Wave Signal of a Core Collapse Supernova Explosion of a 15 M$_\odot$ Star}% Force line breaks with \\

\author{Konstantin N. Yakunin$^{1,2,3}$,  Anthony Mezzacappa$^{1,2}$, Pedro Marronetti$^4$, Eric J. Lentz$^{1,2,3,5}$,
        Stephen W. Bruenn$^6$,\\
        W. Raphael Hix$^{1,3}$, O. E. Bronson Messer$^{1,4,7}$, Eirik Endeve$^{1,2,8}$,
        John M. Blondin$^9$, 
        and J. Austin Harris$^{10}$}

\affiliation{$^1$Department of Physics and Astronomy, University of Tennessee, Knoxville, TN 37996-1200, USA}
\affiliation{$^2$Joint Institute for Computational Sciences, Oak Ridge National Laboratory, P.O. Box 2008, Oak Ridge, TN 37831-6354, USA}
\affiliation{$^3$Physics Division, Oak Ridge National Laboratory, P.O. Box 2008, Oak Ridge, TN 37831-6354, USA}
\affiliation{$^4$Physics Division, National Science Foundation, Arlington, VA 22207 USA}
\affiliation{$^6$Department of Physics, Florida Atlantic University, 777 Glades Road, Boca Raton, FL 33431-0991, USA}
\affiliation{$^5$Joint Institute for Nuclear Physics and its Applications, Oak Ridge National Laboratory, P.O. Box 2008, Oak Ridge, TN 37831-6374, USA}
\affiliation{$^7$National Center for Computational Sciences, Oak Ridge National Laboratory, P.O. Box 2008, Oak Ridge, TN 37831-6164, USA}
\affiliation{$^8$Computer Science and Mathematics Division, Oak Ridge National Laboratory, P.O.Box 2008, Oak Ridge, TN 37831-6164, USA}
\affiliation{$^9$Department of Physics, North Carolina State University,  Raleigh, NC 27695-8202, USA}
\affiliation{$^{10}$Nuclear Science Division, Lawrence Berkeley National Laboratory, Berkeley, CA 94720, US}

\email{kyakunin@utk.edu}

\begin{abstract}
%The current abstract contains 1057 symbols. It has to be reduced to 600 symbols.
In this Letter, we report on the gravitational wave signal computed in the context of an {\em ab initio},  three-dimensional  
simulation of a core collapse supernova explosion, beginning with a 15\msun star and using state-of-the-art weak interactions. The simulation was 
performed with our neutrino hydrodynamics code \chimera. We discuss the potential for detection of our predicted gravitational signal by the 
current generation of gravitational wave detectors.
\end{abstract}
% \keywords{supernovae: general --- gravitational waves --- neutrinos}

%\pacs{Valid PACS appear here}% PACS, the Physics and Astronomy Classification Scheme.

%\keywords{Suggested keywords}%Use showkeys class option if keyword display desired

\maketitle

\emph{Introduction.}--
The first direct detection of gravitational wave (GW) signals from binary-black-hole mergers \cite{GW150914, GW151226} opened a new era in observational astronomy. This has set the stage to prepare, even more fervently, for future detections, especially of other of the primary sources of GWs, among them core-collapse supernovae (CCSNe). CCSNe are physics rich with many physical processes operating in conjunction to produce a supernova. Supernova models are, therefore, innately complex. In the case of a Galactic event, a GW detection is possible \cite{Gossan16}. Such a detection, along with a detection of the SN neutrinos,
would provide {\em direct}  information about these processes and the SN \textquoteleft central engine\textquoteright, in turn allowing us to validate our models. Moreover, in conjunction with detailed GW signals, our models would provide insight into the nature and role of multidimensional fluid instabilities in the proto-neutron star (PNS) and supernova core, the rotation of the stellar core, and the structure of the remnant PNS, as well as the PNS high-density nuclear equation of state (EoS), with implications for fundamental nuclear physics -- e.g., nuclear force models.

Many studies of GW emission in core collapse supernovae (CCSNe) based on a variety of 2D/3D CCSN models were performed in the past \cite{Marek09, Murphy09, Ott09, Ott12, Muller13, Kotake09, Kotake13, Scheidegger10, Fryer11, Abdikamalov14, Kuroda14, Kuroda16a}, including our studies \cite{Yakunin10, Yakunin15}. Progress on multidimensional CCSN modeling has been arguably exponential, in light of the increasingly powerful computational resources available to modelers, culminating in the recent 3D modeling efforts of a number of groups \cite{Kuroda12, Couch13, Hanke13, Ott13, Tamborra13, Nakamura14, Takiwaki14, Tamborra14, Couch15, Foglizzo15, Lentz15, Melson15a, Melson15b, Muller15, Roberts16, Takiwaki16}. A subset of these have been {\em ab initio} simulations with full physics (i.e., general relativistic with a complete set of neutrino interactions) \cite{Muller15, Melson15a, Melson15b, Lentz15}. In turn, a subset of the latter have reported on explosions \cite{Melson15a, Melson15b, Lentz15}.

In this \emph{Letter}, we report on our GW signal predictions based on an {\em ab initio}, general relativistic,
multi-physics, 3D simulation of a CCSN of a 15\msun progenitor using state-of-the-art weak interactions. 
Our signal predictions are based on the simulation data detailed by \citet{Lentz15}.
(See also \citet{Melson15a} for the case of a low-mass progenitor and \citet{Melson15b} for the case of a massive progenitor but with modified neutrino scattering cross sections.) 
Because this simulation includes all of the relevant physics, albeit in some instances with some level of approximation (e.g., ray-by-ray neutrino transport), and because of the simulation outcome, the computed GW signal based on this simulation data is unique, with implications for the prediction of the GW signal 
%. 
%The main characteristics of the waveform, such as its 
amplitude as a function of time (for both polarizations), its frequency distribution and evolution, and the energy emitted in the form of GWs.
%, are better determined than they are in non-exploding (for models that should explode) {\em ab initio}, or in parameterized explosion, models.

\emph{Method and initial models}.-- 
Our GW analysis is based on the data generated in the 2D and 3D core collapse supernova simulations performed by \citet{Lentz15}. The simulations were both initiated from the 15\msun progenitor of \citet{Woosley07} and were carried out with the \chimera code, which includes multigroup flux-limited diffusion neutrino transport with a state-of-the-art set of weak interactions and an effective gravitational potential that incorporates the general relativistic monopole and commensurate corrections (e.g., gravitational redshift) to the neutrino transport \cite{Bruenn16}.
The 3D computational grid comprised 540($r$)$\times$180($\theta$)$\times$180($\phi$) zones equally distributed in the $\phi$-direction only. The $\phi$-resolution was uniformly $2^\circ$. The $\theta$-resolution in the 2D model was uniformly $0.7^\circ$. The $\theta$-resolution in the 3D model varied from $ 2/3^\circ$ near the equator to $8.5^\circ$ near the poles. The radial resolution in both simulations varied according to conditions of the moving grid and  reached  0.1~km inside the PNS.  In both simulations, we employed two equations of states (EoS):  \citet{Lattimer91} (incompressibility $K = 220$~MeV) for $\rho > 10^{11}$~\gcc and an enhanced version of the \citet{Cooperstein85} EoS for $\rho < 10^{11}$~\gcc. In outer regions we employed a 14-species $\alpha$-network \cite{Hix99}. The models were evolved in 1D during collapse and through bounce. At 1.3 ms after bounce random density perturbations of 0.1\% were applied to the matter between 10--30 km.

We compare the GW signals of these two models: C15-2D and C15-3D. 
We employ the quadrupole approximation for extracting the GW signals from the mass motions, using the expressions detailed in \cite{Yakunin15}.
To isolate the impact of dimensionality, all comparisons of the GW emissions in C15-2D and C15-3D have been performed using the same evolution time frame, which is dictated by our 3D run (0--450~ms), and the same sampling interval ($\Delta t = 0.2$~ms). While this is not particularly important for comparisons of the waveforms as a function of time, 
a comparison of the GW spectra is not possible unless the time frame over which the spectra are computed is kept the same.

\begin{figure}[ht!]
\centering
\includegraphics[width=80mm]{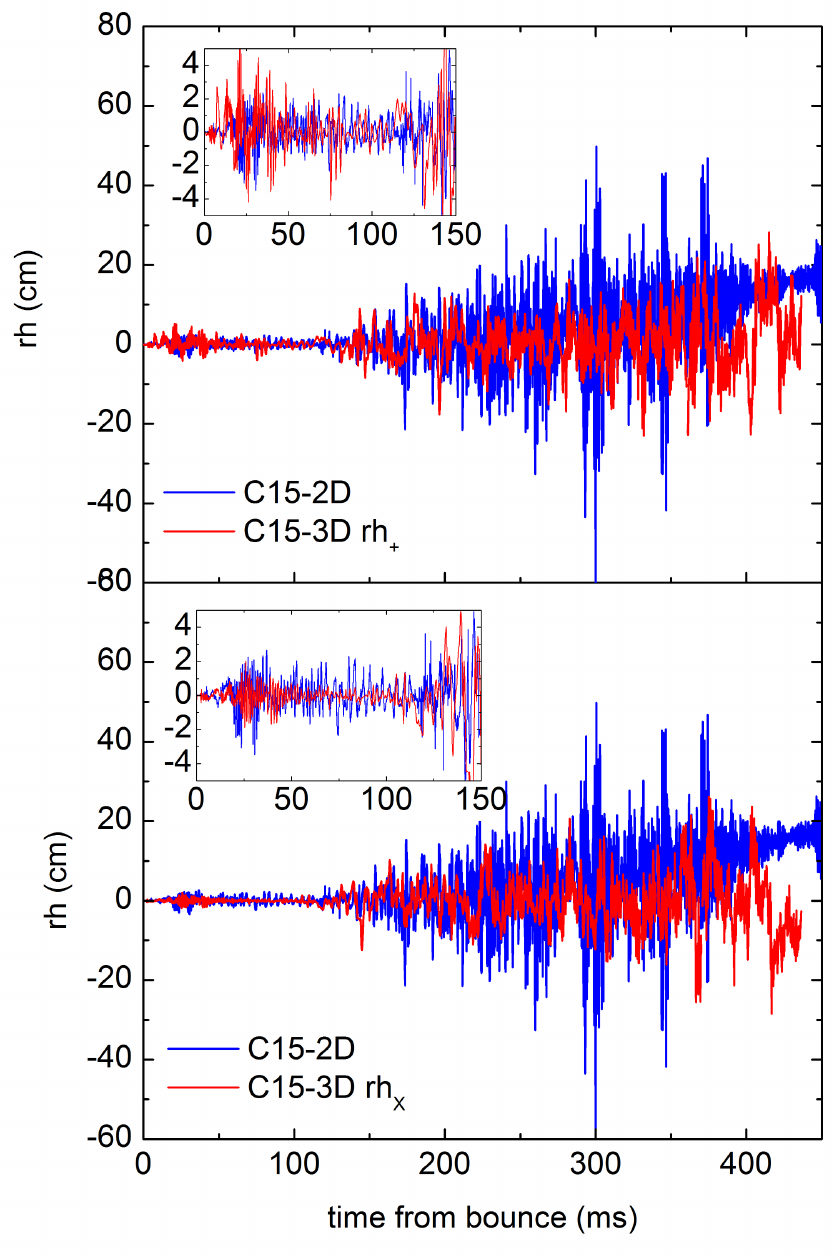}
\caption{
\emph{Top:} The $rh_+$ components of the GW signals produced in the C15-2D and C15-3D models. 
\emph{Bottom:} The $rh_\times$ component of the GW signal produce in the C15-3D 
and the $rh_+$ component of the GW signal from the C15-2D for comparison. 
Both components seen by an equatorial observer. 
\emph{Inset:} The first 150 ms of gravitational waveforms produced in the C15-2D and C15-3D models.
 }
\label{fig:FullSignal}
\end{figure}

\emph{Results.} --
Direct comparison of the GW signal amplitudes from models C15-2D and C15-3D (Fig.~\ref{fig:FullSignal}) should be performed with cognizance of the fact that 3D admits more degrees of freedom -- specifically, that 3D admits two GW polarizations ($+,\times$) whereas 2D admits only one ($+$). \citet{Gossan16} emphasized that the
availability of two independent polarizations in 3D increases the chances of detection by 40\%. %check?

\emph{The prompt convection phase of the GW signal.}--
The early GW signal is produced by Ledoux convection inside the PNS along with matter perturbed behind the quickly expanding shock. It lasts for the first 70--80~ms after bounce. Fig.~\ref{fig:FullSignal} shows that the 3D $rh_+$ signal has generally larger amplitude relative to the 2D signal. However, the difference likely does not arise entirely from the change in dimensionality. As was pointed out in \citet{Yakunin15}, this phase of the signal is very sensitive to model parameters and grid resolution. Usage of a constant-$\mu$ grid for the 3D model (a practical necessity for this model given its computational cost) and a constant-$\theta$ grid for the 2D model likely contributes to the difference. 
3D runs to study the effect of resolution are planned for the near future.
Resolution aside, it is clear that an early GW signal phase is present in 3D, in both polarizations, and possibly more vigorous in the $+$ polarization relative to the 2D case. 

\emph{The quiescent phase of the GW signal.}--
The quiescent phase -- a transition phase between the early and strong signal phases -- is coincident (80--120~ms after bounce) in the two models. Moreover, the signal during this phase is similar in magnitude in each model. As we will discuss, the fundamental difference between the C15-2D and C15-3D signals does not arise until neutrino-driven convection and the standing accretion shock instability (SASI) \cite{Blondin03} develop. The quiescent phase is particularly evident in the plot of $rh_\times(t)$  in Figure \ref{fig:FullSignal}.

\begin{figure}[ht!]
\centering
\includegraphics[scale=0.45]{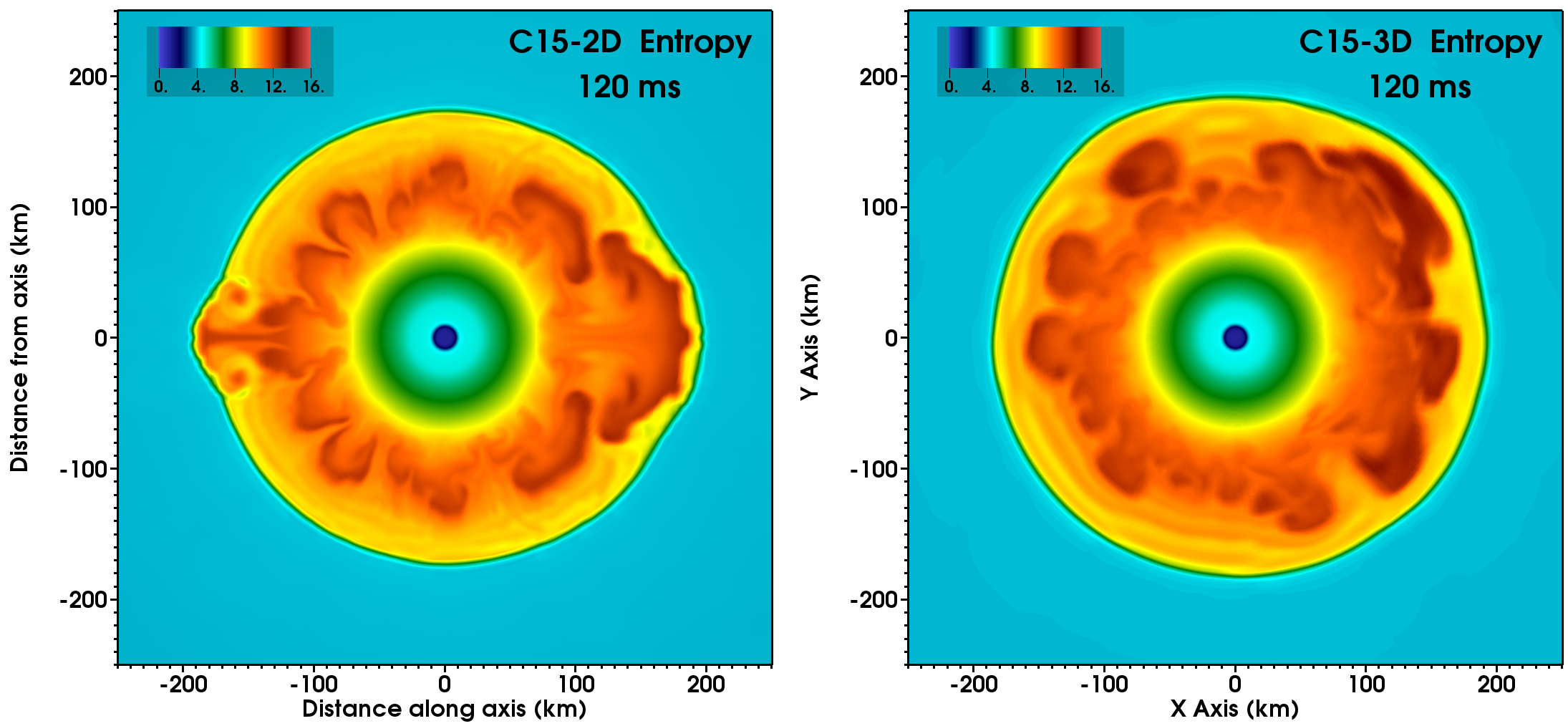}
\caption{Entropy distributions for the C15-2D model (\emph{left}) and for the equatorial slice of the C15-3D model (\emph{right}) at 120 ms after bounce. }
\label{fig:EntropySlice}
\end{figure}

\emph{The neutrino-driven convection/SASI (i.e., strong) phase of the GW signal.}--
The strongest phase of the GW signal starts at $\sim$120~ms for both our 2D and 3D simulations. This is not surprising given that the shock is still quasi-spherical at that time (see Fig.~\ref{fig:EntropySlice}), and the angle-averaged shock trajectories in 2D and 3D still follow one another rather closely \cite{Lentz15}. However, after $\sim$150~ms the shock behavior is very different in the two models. In the 2D model, we observe global oscillations of the shock due to the SASI \cite{Bruenn16}, but the SASI is less pronounced in 3D during the initial $\sim$450~ms window considered here \cite{Lentz15}. Global SASI oscillations in the 2D case induce the formation of a few, massive accretion funnels. 
When these funnels impinge on the surface of the PNS, they perturb the PNS and generate the high-frequency, large-amplitude excursions evident in $rh_+$.
In the 3D case, there are a larger number of funnels from neutrino-driven convection, each accreting less mass than those in 2D, that impinge on the PNS to induce the same behavior in both $rh_+$ and $rh_\times$. In performing a comparison of the amplitudes of $rh_+$ across the 2D and 3D cases, it is important to keep in mind the presence of two polarizations in the 3D case. The overall structure of the GW signals, and their relationship to the phenomenology of the stellar core, is similar, and consistent (in the sense that there is no physical argument as to why we should expect any of the phases associated with the 2D GW signal to be absent in 3D), across the 2D and 3D cases. A comparison of the signals in frequency, discussed below, supports this. Both the spectra and the total energy emitted are based on the {\em complete} GW signals -- in the 2D case, on the $+$ polarization, and in the 3D case, on {\em both} the $+$ and $\times$ polarizations. In 3D and during the first $\sim$450~ms, we still see three phases of GW emission. The prompt convection phase is very similar in timing and amplitude to the 2D case. The quiescent phase is present, and starts and stops at the same post-bounce times relative to the 2D case. The third phase begins at the same post-bounce time relative to the 2D case, exhibits the same qualitative behavior, as neutrino-driven convection and the SASI develop and induce downflows onto the PNS. The fourth, explosion, phase has not yet been observed in the 3D case.

\emph{The explosion phase of the GW signal.}--
Rapid shock expansion at the onset of explosion in the 3D case is delayed by approximately 100~ms relative to its 2D counterpart \cite{Lentz15}. Consequently, during the $\sim$450~ms considered here, we do not observe the low-frequency tail of $rh_+$ associated with prolate or oblate explosion. 
In the 2D case, the tail is evident in the final 100~ms of available data, and given the $\sim$100~ms delay of explosion in 3D, another $\sim$100~ms beyond the end of the 3D simulation presented here would be needed to see evidence of the expected tail.
However, we do expect the magnitude of the tail in the 3D case to be smaller. 
The high-entropy bubble that initiates the outward acceleration of the shock \cite{Lentz15} in 3D contains a smaller mass fraction than that in the 2D case, where the high-entropy bubble takes up much of the volume behind the shock.

\begin{figure}[ht!]
\centering
\includegraphics[width=80mm]{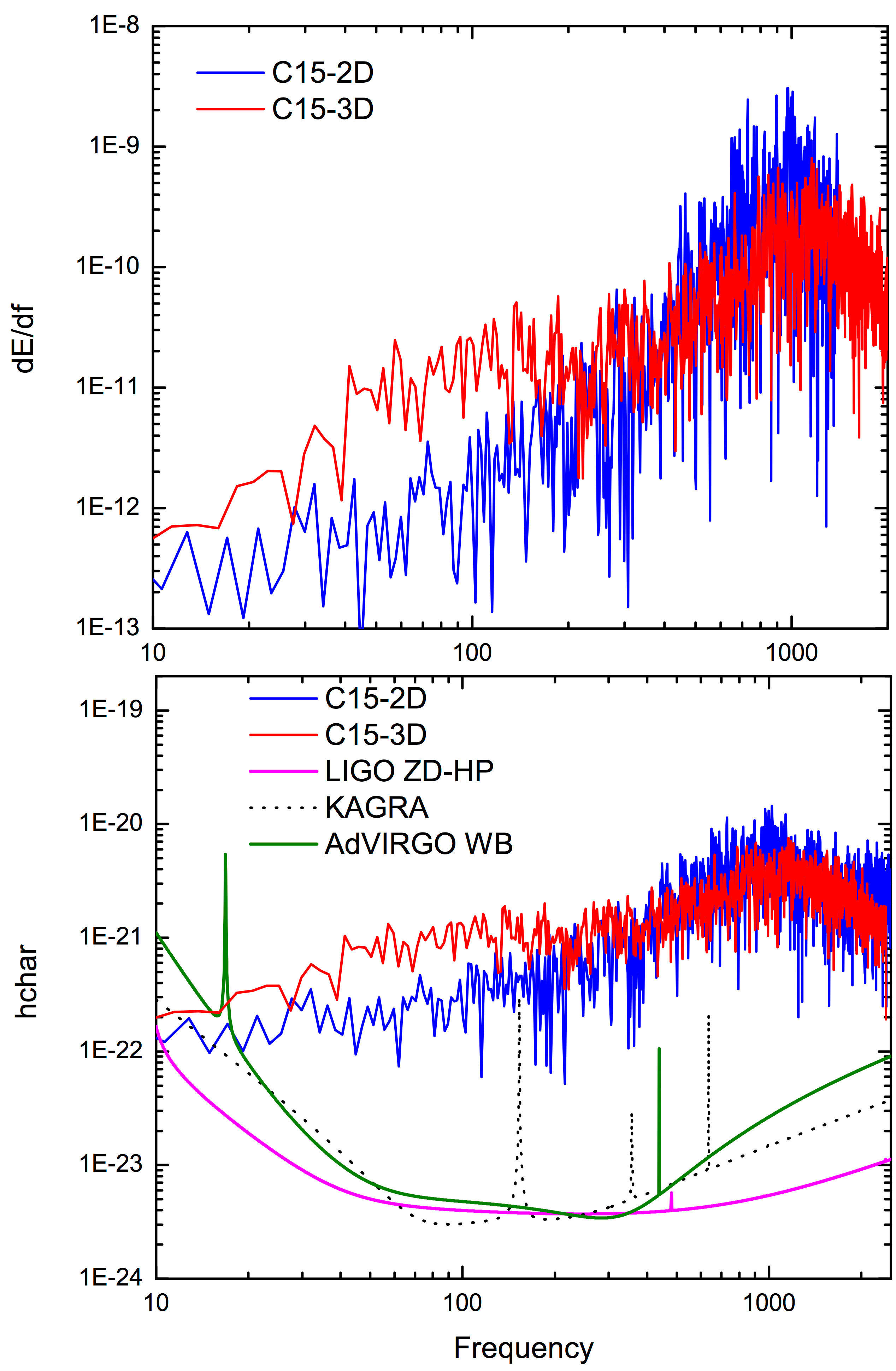}
\caption{Frequency analysis of the GW signals from C15-2D and C15-3D models. \emph{Top:} Spectral energy density distributions for 2D and 3D models. \emph{Bottom:} Characteristic spectral strain  $h_\textrm{char}(f)f^{-1/2}$ of 2D and 3D models at a distance of 10~kpc compared with the design noise levels $\sqrt{S(f)}$ of Advanced LIGO in the broadband zero-detuning high-power mode (aLIGO ZD-HP), KAGRA, and Advanced Virgo in wideband mode (AdV WB).}
\label{fig:FFT}
\end{figure}

\emph{The GW energy spectra and the integrated GW energy emission with time.}--
Figure \ref{fig:FFT} plots the decomposition of our C15-2D and C15-3D GW signals. Both are computed at the same post-bounce time of 450 ms. Qualitatively, the GW spectra are very similar. 
A frequency shift in the energy spectra, from high (C15-2D) to low (C15-3D) frequencies, is evident in the time-integrated spectral energy distribution dE/df.  
This results from the lack of axisymmetry ($l=2$, $m=0$ only) in 3D that allows a transfer of convective energy into multiple $l=2$ modes ($l=2$, $m=-2, ... ,+2$) that are absent in 2D.
Figure~\ref{fig:FFT} also shows the angle-averaged characteristic GW strain spectra $h_\textrm{char}(f)$ \cite{Flanagan98} of our 2D and 3D models, along with the broadband design noise levels of advanced-generation GW interferometers, assuming a source distance of 10~kpc. Most of the detectable emission is within 20--2500~Hz and at essentially the same level of $\sim$1-4 of $10^{-21}$~Hz$^{-1/2}$. A Galactic event (at 10~kpc) appears to be well detectable by upcoming detectors.

\begin{figure}[ht!]
\centering
\includegraphics[width=90mm]{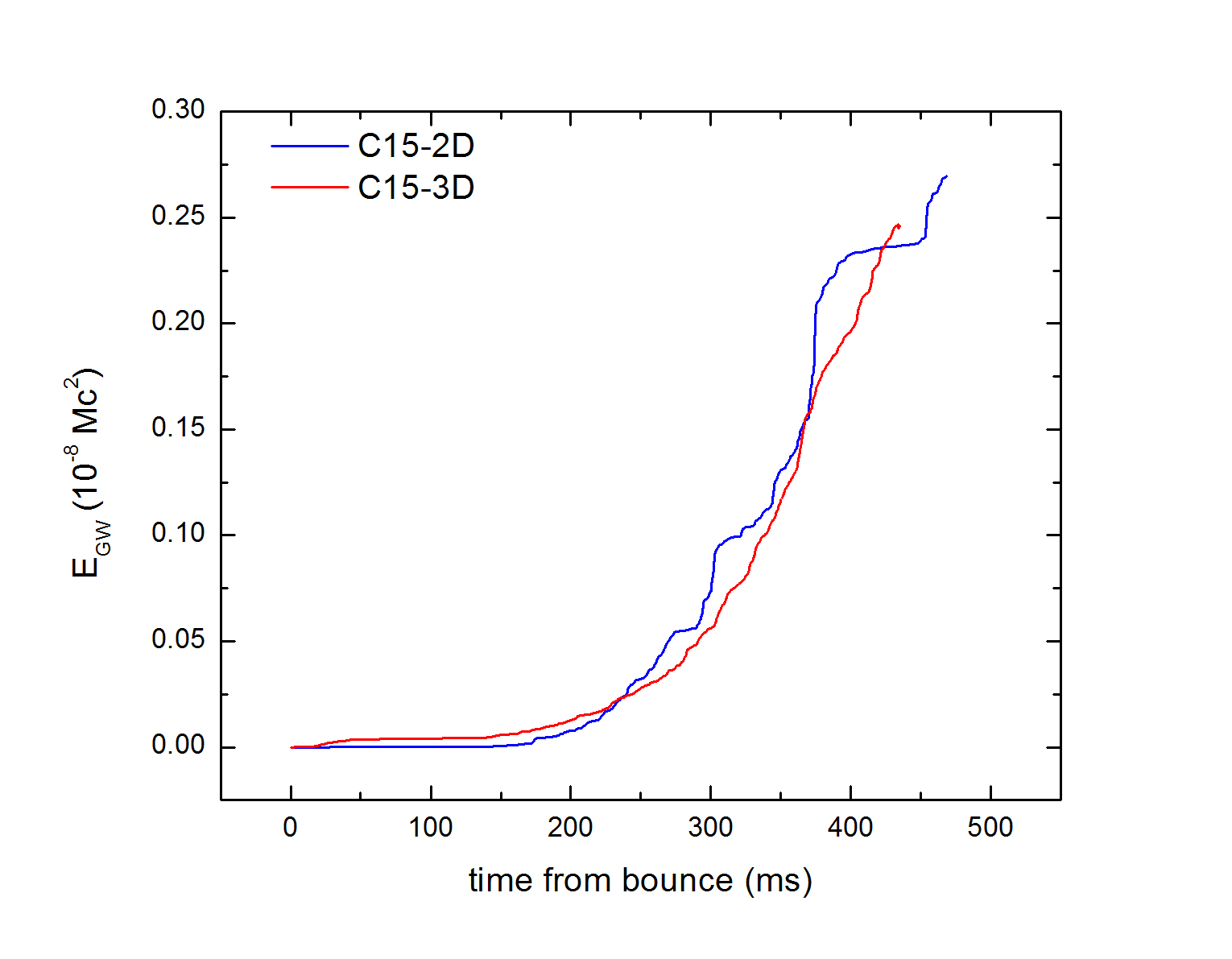}
\caption{Energy emitted in the form of GWs during the first 440 ms of CCSN explosion for the C15-2D, and C15-3D models. The step-like behavior of $E_\textrm{GW}$ in the C15-2D model (see, for instance, 300~ms and 400~ms) reflects the evolution of the single dominant accretion downflow in this model, in contrast to the multiple downflows in the 3D model.}
\label{fig:Egw}
\end{figure}

The total, frequency-integrated GW energy emitted is plotted in Figure~\ref{fig:Egw}. The emitted GW energy in C15-2D and C15-3D follow one another closely. However, the fundamentally different character of the mass accretion onto the PNS between the 2D and 3D cases is imprinted here. The jumps in the emitted GW energy clearly seen in our 2D model correspond to sudden increases in the accretion rate. In the 2D case, mass accretion is mediated by a few, massive funnels. The addition, or loss, of such a funnel would be accompanied by a significant change in the mass accretion rate. On the contrary, the jumps in the emitted GW energy are absent in the 3D case, with this energy growing smoothly with post-bounce time. 
In the 3D case, mass accretion onto the PNS is mediated by numerous, 
lower-mass accretion funnels. 
The addition, or loss, of such a funnel would not result in a significant change in the mass accretion rate. 
From both GW spectra and frequency-integrated GW energy emission, we see that the numerous downflows we observe in our 3D model produce a response of the PNS similar to the fewer, more massive downflows observed in our 2D model. Thus, one may conclude that the characteristic frequency of the GW signals depend on the internal properties of the PNS (e.g., its high-density nuclear equation of state and the associated radius, density profile, etc.), whose evolution is similar in C15-2D and C15-3D.

\emph{Summary, Discussion, and Outlook.}--
Our {\em ab initio}, multi-physics, 3D simulation of a CCSN explosion allowed us to compute the detailed time dependence of the GW signals for both the $h_+$ and $h_\times$ polarizations. For the time window considered here, which is approximately the first half second of evolution after stellar core bounce, we provide the corresponding spectral decomposition of the total signal, as well as the total energy emitted in GWs (from matter) as a function of post-bounce time. Quantitatively, the 3D signals differ from the signals obtained in our 2D counterpart model 
for the reasons covered above,
but they confirm the existence of the first 3 phases of the GW signal accessible in this study: a prompt convection phase followed by a quiescent phase followed by the dominant GW emission phase from neutrino-driven convection and the SASI. The final, explosion phase is not accessible at this time given the simulation presented here covers only the first half second after bounce. The results presented here are qualitatively consistent with the prediction of a 4-phase signal first presented by \citet{Murphy09}. Clearly, our understanding of the GW signals from (neutrino-driven) core collapse supernovae -- the details of the signals and their association with the underlying CCSN phenomenology -- is maturing.

Equally important, the predicted signals presented here were shown to be detectable by LIGO and other extant GW observatories for a Galactic CCSN event. For a detailed study of the detectability of such an event, we refer the reader to the work by \citet{Gossan16}.

Recently, \citet{Andresen16} documented their predictions for the GW signatures from several of their 3D models, although for different progenitors than the ones considered here. Three of the four models presented by \citet{Andresen16} do not explode. One model does explode when modifications are made to the axial vector coupling constant in the neutral-current scattering cross sections. In all four cases, these authors find no significant GW production for the first $\sim$175~ms after bounce. 
The differences between our predictions for the early GW signal and the predictions of the Garching group will need to be explored further.
However, for non-rotating (spherical) progenitors, we do not expect to see significant differences in the 2D and 3D cases, 
which is what we observe (see \citet{Yakunin15} for the 2D case). The conditions in the inner regions of the PNS at these earliest 
times after bounce simply do not differ significantly as we move from 2D to 3D. In contrast, the 3D predictions of \citet{Andresen16} differ from the
Garching group's predictions in the 2D case \cite{Muller13} for the same progenitor masses within the same set of progenitors (e.g., WHW02 vs. WH07).
Of course, it is not clear how productive a comparison of the early signal obtained by different groups is in the case of non-rotating (spherical)
progenitors. A much more robust early signal will be obtained in the context of first-principles simulations with rotating progenitors, 
which will produce strong signals at bounce. Moreover, the bounce signals will be less sensitive to simulation details (numerical methods, 
grid resolution, etc.) and, rather, will depend on the physical initial conditions assumed.
At later times after bounce, during the dominant phase of GW emission associated with neutrino-driven convection and the SASI, we do see a reduction of the amplitude of the strain in the 3D case, relative to the 2D case, for the $+$ polarization, but in 3D the emission is shared between the $+$ and $\times$ polarizations, making a comparison difficult.
We do not see significant differences in the GW energy emitted as a function of time between the two cases, and, for the first $\sim$450~ms, the 3D GW spectrum exhibits a similar structure relative to the 2D case, including a peak in the spectrum at $\sim$1000~Hz. 
%In summary, we do not see the ``profound'' differences documented by \citet{Andresen16} in going from 2D to 3D.

While the GW signal predictions presented here are based on {\em ab initio} models that exhibit a noteworthy level of realism, future models can and should develop in obvious ways: (1) While the use of the ``ray-by-ray'' neutrino transport approximation may be a better approximation in 3D than in 2D \cite{Skinner15}, definitive 3D models will require 3D neutrino transport. (2) The use of the GR monopole correction to the Newtonian self-gravitational potential should be replaced by a more sophisticated treatment of GR, such as the Conformally Flat Approximation (CFA) \cite{Mueller10}, or a full BSSNOK treatment \cite{Kuroda16b}. (3) To fully resolve turbulent cascades in 3D CCSN simulations, which is relevant for GW predictions, one requires angular grid resolutions of less than $1^\circ$ and radial resolutions inside the PNS of less then 0.1~km \cite{Radice16, Weinberg08}. 
(4) Future 3D CCSN models will need to begin with 3D progenitor models. 
Recent studies have demonstrated that the impact of improved initial conditions on CCSN models and their predicted outcomes is potentially significant \cite{Couch15}.
\\
\par
This research was supported by the U.S. Department of Energy Offices of Nuclear Physics and Advanced Scientific Computing Research, the NASA Astrophysics Theory and Fundamental Physics Program (grants NNH08AH71I and NNH11AQ72I), and the National Science Foundation PetaApps Program (grants OCI-0749242, OCI-0749204, and OCI-0749248) and Gravitational Physics Program (grant GP-1505933). 
This research was also supported by an award of computer time provided by the Innovative and Novel Computational Impact on Theory and Experiment (INCITE) Program at the Oak Ridge Leadership Computing Facility (OLCF) and at the Argonne Leadership Computing Facility (ALCF), which are DOE Office of Science User Facilities supported  under contracts DE-AC05-00OR22725 and DE-AC02-06CH11357, respectively.
P. M. is supported by the National Science Foundation through its employee IR/D program. The opinions and conclusions expressed herein are those of the authors and do not represent the National Science Foundation. 

\bibliographystyle{apsrev4-1} 

\bibliography{GW-3D} % expects file "GW-3D.bib"

\end{document}